\documentclass[oneside,12pt,a4paper]{article}

\usepackage[ansinew]{inputenc}
\usepackage{booktabs, tabularx}
\usepackage{epsfig}
\usepackage{float}
\usepackage{picins}
\usepackage{amsmath}
\usepackage{amsfonts}
\usepackage{amssymb}
\usepackage{accents}
\usepackage{enumerate}
\usepackage{afterpage}
\usepackage{textcomp}
\usepackage{setspace}
\floatstyle{plain}

\begin{document}

\begin{center}
  {\Large \bf  Essentially ML ASN-Minimax double sampling plans}\\[0.5cm]
   {\sc Eno Vangjeli}\\[0.3cm]
\end{center}
\vspace*{0.1cm}  {\small }
{\small{\bf{Abstract:}} Subject of this paper is ASN-Minimax (AM) double sampling plans by variables for a normally distributed quality characteristic with unknown standard deviation and two-sided specification limits. Based on the estimator $p^*$ of the fraction defective $p$, which is essentially the Maximum-Likelihood (ML) estimator, AM-double sampling plans are calculated by using the random variables $p^*_1$ and $p_p^*$ relating to the first and pooled samples, respectively. Given $p_1$, $p_2$, $\alpha$, and $\beta$, no other AM-double sampling plans based on the same estimator feature a lower maximum of the average sample number (ASN) while fulfilling the classical two-point condition on the corresponding operation characteristic (OC).}\\
\begin{center}
1. {\sc Introduction}
\end{center}
\vspace*{2mm}
When carrying out sampling inspection for a normally distributed characteristic $X\sim N(\mu, \sigma)$, $\sigma>0$ the following four cases arise:
\begin{enumerate}[(i)]
\item
One-sided specification limit, $\sigma$ known
\item
Two-sided specification limits, $\sigma$ known
\item
One-sided specification limit, $\sigma$ unknown
\item
Two-sided specification limits, $\sigma$ unknown.
\end{enumerate}
In this paper, we deal with ASN-Minimax (AM) double sampling plans for case (iv). Let $L$ be a lower and $U$ an upper  specification limit to $X$. The fraction defective function $p(\mu, \sigma)$ is defined as:
\begin{equation}
p(\mu, \sigma):=P(X<L)+P(X>U)=\Phi\left({L-\mu\over \sigma}\right)+\Phi\left({\mu-U\over \sigma}\right),
\end{equation}
where $\Phi$ denotes the standard normal distribution function. Note, $p(\mu, \sigma)$ is a three-dimensional function. For different levels of $p$, corresponding iso-p-lines arise symmetrically to $\displaystyle \mu_0=\frac{L+U}{2}$ on the $\mu$-$\sigma$-plane. A figure containing different iso-p-lines can be found in {\sc Bruhn-Suhr} and {\sc Krumbholz} (1990).
Given a large-sized lot, a single sample $X_1,...,X_n$, $(n>3)$ with
\begin{equation*}
\overline{X}={1\over n}\sum_{i=1}^{n}X_i, \quad  S^2={1\over n-1}\sum_{i=1}^{n}(X_i-\overline{X})^2,
\end{equation*}
an acceptable quality level $p_1$, a rejectable quality level $p_2$ and levels $\alpha$ and $\beta$ of Type-I and Type-II error, respectively, {\sc Bruhn-Suhr} and {\sc Krumbholz} (1990) develop single sampling plans based on the essentially Maximum Likelihood (ML) estimator
\begin{equation}
p^*=p(\overline{X}, S)=\Phi\left({L-\overline{X}\over S}\right)+\Phi\left({\overline{X}-U\over S}\right).
\end{equation}
The lot is accepted within the single plan $(n, ~k)$, if  $p^*\leq k$.\\
With the help of the operation characteristic (OC) of single sampling plans, {\sc Vangjeli} (2011) develops AM-double sampling plans $\lambda^*_1$ based on the independent random variables  $p^*_1$ and $p^*_2$, which relate to the first and second samples, respectively. Given $p_1$, $p_2$, $\alpha$, and $\beta$, the AM-double sampling plan fulfills the classical two-points-condition on the OC and features the lowest maximum of the average sample number (ASN). $\lambda^*_1$ is computed in a similar fashion to the corresponding single sampling plan $(n, ~k)$ by using its one-sided approximation AM-double sampling plan $\widetilde{\lambda}_1$, which is based on information obtained only from the second sample in the second stage. A double sampling plan consisting of two independent consecutive samples needs a larger sample size to fulfill the classical two-points-condition on its OC than the corresponding double sampling plan defined by taking into account information from both samples in the second stage.\\
In this paper, we introduce the AM-double sampling plan $\lambda^*_2$ based on the random variables  $p^*_1$ and $p_p^*$. Using the random variable $p_p^*$, which contains information from both samples in the second stage, the OC of an arbitrary double sampling plan $\lambda_2$ becomes more complex than the OC of the corresponding double sampling plan $\lambda_1$. The probability for accepting the lot after the inspection of the first sample is analogously to $\lambda_1$ a single-sampling-plan-OC. Thus, in the next section some preliminaries regarding the single-sampling-plan-OC, as well as notation and definitions concerning the double sampling plan $\lambda_2$ are introduced. The increased complexity of $\lambda_2$-OC compared to $\lambda_1$-OC is found in the probability for accepting the lot after the inspection of the second sample. The derivation of this probability is described in Section 3. The AM-double sampling plan $\lambda^*_2$ is computed analogously to $\lambda^*_1$ by using the corresponding one-sided approximation AM-double sampling plan $\widetilde{\lambda}_2$. A comparison between $\lambda^*_1$ and $\lambda^*_2$ is presented in Section 4.\\
\begin{center}
2. {\sc Preliminaries}
\end{center}
\vspace*{2mm}
Before introducing the notation and definitions for deriving the double-sampling-plan-OC, we first note a well-known issue from single sampling. Let
\begin{equation}
 L_{(n, ~k)}(\mu, \sigma)=P(p^*\leq k)
\end{equation}
denote the OC for the single plan $(n, ~k)$ and let $g_r$ be the density function of the $\chi^2$ distribution with $r$ degrees of freedom.\\\\
{\bf Theorem 1:} \quad It holds that:
\begin{eqnarray}
&& L_{(n, ~k)}(\mu, \sigma)=\int_0^B\biggl\{  \Phi\left({\sqrt{n}\over\sigma}\left(\mu\left(\sigma \sqrt{{t \over n-1}}, k \right)-\mu\right)\right)\nonumber\\
&&-\Phi\left({\sqrt{n}\over\sigma}\left(\dot{\mu}\left(\sigma \sqrt{{t \over n-1}}, k \right)-\mu\right)\right) \biggr\}g_{n-1}(t)dt
\end{eqnarray}
with
\begin{equation*}
B={(n-1)(L-U)^2 \over 4\sigma^2\left(\Phi^{-1}\displaystyle\left({k\over2}\right)\right)^2} \quad \text{and} \quad \dot{\mu}(\sigma, p)=L+U-\mu(\sigma, p).
\end{equation*}
For the proof of Theorem 1, {\sc Bruhn-Suhr} and {\sc Krumbholz} (1990) use the fact that for a given $\accentset{\circ}{p} $ $(0<\accentset{\circ}{p}<1)$ and $\accentset{\circ}{\sigma}>0$,
\begin{equation}
M(\accentset{\circ}{\sigma}, \accentset{\circ}{p}):=\{ \mu \in \mathbb{R}~ |~ p(\accentset{\circ}{\sigma}, \mu)\leq \accentset{\circ}{p} \}
\end{equation}
is equivalent to
\begin{equation}
M(\accentset{\circ}{\sigma}, \accentset{\circ}{p})=
\begin{cases}
[\dot{\mu}(\accentset{\circ}{\sigma}, \accentset{\circ}{p}), \mu(\accentset{\circ}{\sigma}, \accentset{\circ}{p})]& \text{if} \quad \accentset{\circ}{\sigma}\leq\sigma_0(\accentset{\circ}{p})\\
\hspace*{1.3cm}\emptyset & \text{otherwise},
\end{cases}
\end{equation}
with
\begin{equation}
\sigma_0(\accentset{\circ}{p})={L-U\over \displaystyle2\Phi^{-1}\left({\accentset{\circ}{p} \over 2}\right)}.~~~ \text{(See Figure 1)}
\end{equation}
\vspace*{-20mm}
\begin{figure}[H]
\begin{center}
   \includegraphics[width=1\columnwidth]{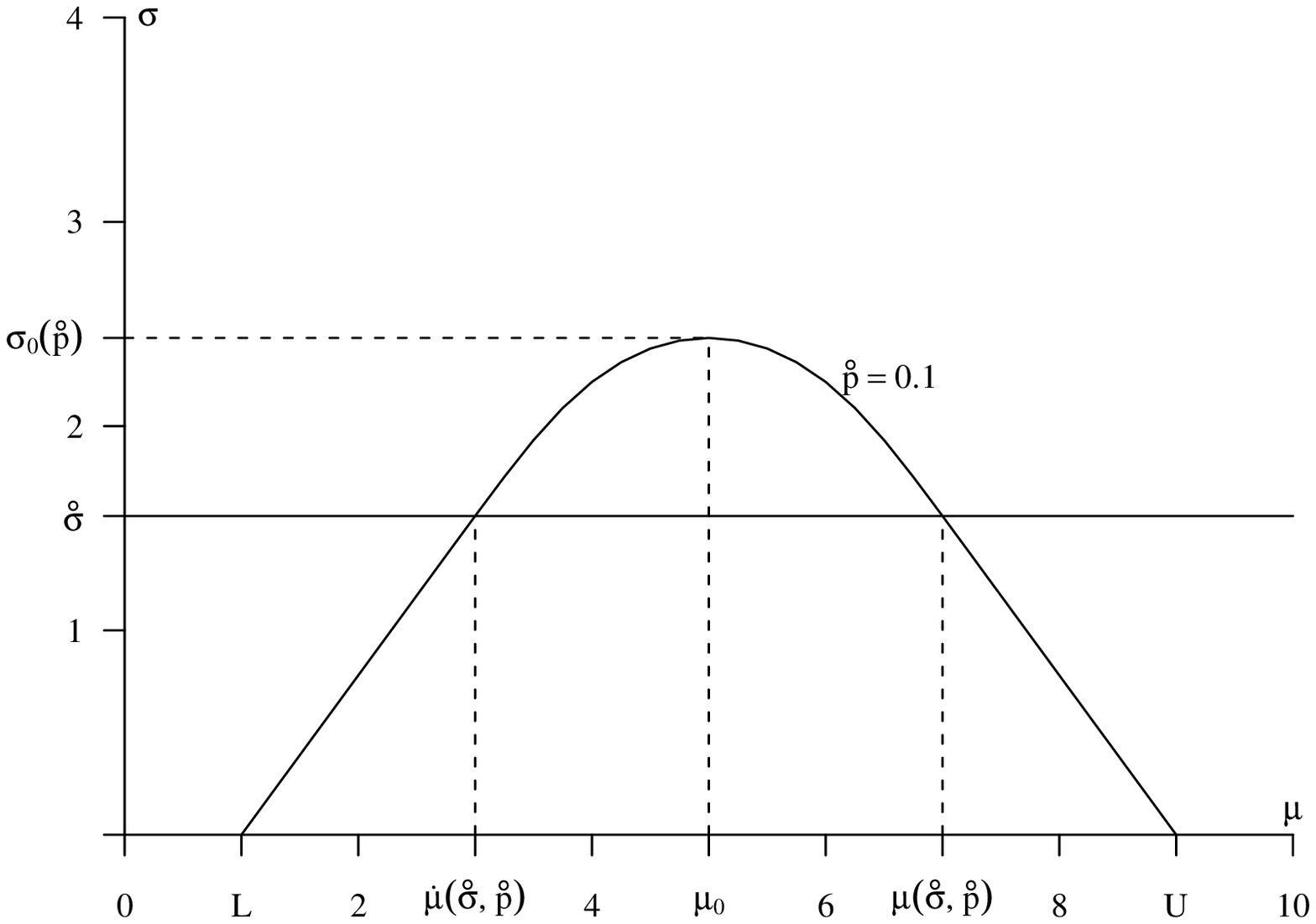}
\end{center}
\end{figure}
\vspace*{-15mm}
\hspace*{-6mm}Figure 1: Iso-p-line for $\accentset{\circ}{p}=0.1$ with $\mu_0=5, ~\sigma_0=2.431827$ and $M(\accentset{\circ}{\sigma}, \accentset{\circ}{p})$ for $\accentset{\circ}{\sigma}=1.560192$\\\\
Now, we turn our attention to the double sampling plan $\lambda_2$. Let $X_1,...,X_{n_1}$ be the first and $X_{n_1+1},...,X_{n_1+n_2}$ the second sample on $X$. Then, define the following notation:
\begin{equation}
\overline{X}_1  =  {1\over n_1}\>\>\sum_{i=1}^{n_1}\>X_i,
\end{equation}
\begin{equation}
S_1^2  =  {1\over n_1-1}\>\>\sum_{i=1}^{n_1}\>(X_i-\overline{X}_1)^2={1\over n_1-1}\>\left(\sum_{i=1}^{n_1}\>X_i^2-n_1\overline{X}_1^2\right),
\end{equation}
\begin{equation}
\overline{X}_2  =  {1\over n_2}\> \sum_{i=n_1+1}^{n_1+n_2}\>X_i,
\end{equation}
\begin{equation}
\stackrel{=}{X} =  {1\over n_1+n_2}\>\>\sum_{i=1}^{n_1+n_2}\>X_i={n_1\>\overline{X}_1+n_2\>\overline{X}_2\over n_1+n_2},\\
\end{equation}
\begin{equation}
S^2  =  {1\over n_1+n_2-1}\>\sum_{i=1}^{n_1+n_2}\>(X_i-\stackrel{=}{X})^2.
\end{equation}
{\bf Definition 1:} \quad The double sampling plan by variables $\lambda_2 = \left(\begin {array}{c c c}
n_1 & k_1 & k_2\\ n_2 & k_3
\end {array}\right) $
with $n_1, n_2\>{\in\mathbb{N}};\>\>n_1, n_2\geq 2;\>\>k_1, k_2, k_3\>{\in\mathbb{R}^+};\>\>k_1\leq
k_2$, is defined as follows:
\begin{enumerate}[(i)]
\item
Observe a first sample of size  $n_1$ and compute $p^*_1=p(\overline{X}_1, S_1)$.\\\\
If $p^*_1 \leq k_1$, accept the lot.\\
If $p^*_1 > k_2$, reject the lot.\\
If $k_1< p^*_1 \leq k_2$, go to (ii).
\item
Observe a second sample of size $n_2$ and compute $p_p^*=p(\stackrel{=}{X}, S)$.\\\\
If $p_p^* \leq k_3$, accept the lot.\\
If $p_p^* > k_3$, reject the lot.\\\\
\end{enumerate}
\vspace*{-.7cm}
The $\lambda_2$-OC is given by
\begin{equation}
L_{\lambda_2} (\mu, \sigma)=P_{(\mu, \sigma)}(A_1)+P_{(\mu, \sigma)}(A_2)
\end{equation}
with
\begin{equation}
A_1=\{p^*_1 \leq k_1\},\>A_2=\{p_p^* \leq k_3,\>k_1< p^*_1\leq k_2\}.
\end{equation}
From (3), (4) and (14) it follows that
\begin{equation}
P_{(\mu, \sigma)}(A_1)=L_{(n_1, ~k_1)}(\mu, \sigma).
\end{equation}
Since $P(A_2):=P_{(\mu, \sigma)}(A_2)$ is more complex, we describe how to determine it in the next section. The $\lambda_2$-ASN is given by
\begin{equation}
N_{\lambda_2}(\mu, \sigma)=n_1+n_2 P_{(\mu, \sigma)}(k_1 < p^*_1 \leq k_2)
\end{equation}
with
\begin{equation*}
P_{(\mu, \sigma)}(k_1 < p^*_1 \leq k_2)=L_{(n_1, k_2)}(\mu, \sigma)-L_{(n_1, k_1)}(\mu, \sigma).
\end{equation*}\\
{\bf Remark 1:} The following analogies between $\lambda_1$ and $\lambda_2$ hold:
\begin{enumerate}[(i)]
\item
$L_{\lambda_2} (\mu, \sigma)$ and $N_{\lambda_2}(\mu, \sigma)$ are not unique functions in $p$, but bands.
\item
Let the symbol $^*$ indicate the AM-double sampling plan. Denoting $\phi_1^*$ as the one-sided AM-approximation for $\lambda^*_1$, {\sc Vangjeli} (2011) shows that there are nonessential differences between $N_{max}(\lambda^*_1)$ and $N_{max}(\phi^*_1)$ \footnote{The examples given in the fourth section confirm this fact for $N_{max}(\lambda^*_2)$ and $N_{max}(\phi^*_2)$}.\\
\end{enumerate}
\begin{center}
3. {\sc The $P(A_2)$}
\end{center}
\vspace*{2mm}
Let
\begin{equation}
P(A_2^{u}):=P_{(\mu, \sigma)}(A_2^{u})=P_{(\mu, \sigma)}(p_p^* \leq k_3,\>p^*_1\leq k_2)
\end{equation}
and
\begin{equation}
P(A_2^{l}):=P_{(\mu, \sigma)}(A_2^{l})=P_{(\mu, \sigma)}(p_p^* \leq k_3,\>p^*_1\leq k_1).
\end{equation}
The probability
\begin{equation}
P(A_2):=P_{(\mu, \sigma)}(A_2)=P_{(\mu, \sigma)}(p_p^* \leq k_3,\>k_1< p^*_1\leq k_2)
\end{equation}
can be written as
\begin{equation}
P(A_2)=P(A_2^{u})-P(A_2^{l}).
\end{equation}
For $i=1, 2$, let
\begin{equation}
Y_i:=\sqrt{n_i}\>{\overline{X}_i-\mu\over \sigma}\>\sim N(0,1)
\end{equation}
and
\begin{equation}
W_i:={n_i-1\over \sigma^2}\>S_i^2\>\sim \chi^2_{n_i-1}.
\end{equation}
{\sc Krumbholz} and {\sc Rohr} (2006) have shown that the following holds:
\begin{equation}
S={\sigma\>\sqrt{(n_1+n_2)\>(W_1+W_2)+(\sqrt{n_2}\>Y_1-\sqrt{n_1}\>Y_2)^2}\over \sqrt{(n_1+n_2-1)(n_1+n_2)}}.
\end{equation}
Along with (21), it can be shown that
\begin{equation}
\stackrel{=}{X} = {\sqrt{n_1}\>\sigma(Y_1+\sqrt{n_1}\>\displaystyle{\mu\over\sigma})+\sqrt{n_2}\>\sigma(Y_2+\sqrt{n_2}\>{\mu\over\sigma})\over n_1+n_2}.
\end{equation}
Due to total probability decomposition and the independence of $\stackrel{=}{X}$ and $S^2$, $P(A_2^{u})$ can be written as:
\begin{eqnarray}
&&\hspace*{-1.15cm}P(A_2^{u})=\int\limits_0^{\infty}\biggl(\int\limits_{-\infty}^{\infty}\>\biggl(\int\limits_{-\infty}^{\infty}\>\biggl(\int\limits_0^{\infty}\>P(A_2^{u}|W_1=w_1,\>Y_1=y_1,\>Y_2=y_2,\>W_2=w_2)\times\nonumber\\ &&\hspace*{0.15cm}\times\>g_{n_{2}-1}(w_2)\>dw_2\biggr)\>\Phi'\>(y_2)\>dy_2\biggr)\>\Phi'\>(y_1)\>dy_1\biggr)\>g_{n_{1}-1}(w_1)\>dw_1.
\end{eqnarray}
It holds that:
\begin{eqnarray}
&&\hspace*{-2.6cm}P(A_2^{u}|W_1=w_1,\>Y_1=y_1,\>Y_2=y_2,\>W_2=w_2)=\nonumber \\
&&\hspace*{3.cm}=P(p(\stackrel{=}{X}, S) \leq k_3,\>p(\overline{X}_1, S_1)\leq k_2).
\end{eqnarray}
From (6) and (24), for $S<\sigma_0(k_3)$, we get:
\begin{equation}
p(\stackrel{=}{X}, S) \leq k_3~~\Leftrightarrow ~~ \dot{\mu}(S, k_3)\leq\> \stackrel{=}{X}\>\leq \mu(S, k_3)~~\Leftrightarrow ~~ C_1\leq Y_2\leq C_2,
\end{equation}
where
\begin{equation}
C_1={(n_1+n_2)\>\dot{\mu}(S, k_3)-(\sigma \sqrt{n_1}\>Y_1+(n_1+n_2)\>\mu)\over \sigma\>\sqrt{n_2}}
\end{equation}
and
\begin{equation}
C_2={(n_1+n_2)\>\mu(S, k_3)-(\sigma \sqrt{n_1}\>Y_1+(n_1+n_2)\>\mu)\over \sigma\>\sqrt{n_2}}.
\end{equation}
From (23) and $S<\sigma_0(k_3)$, it follows that
\begin{equation}
W_2 \leq D,
\end{equation}
where
\begin{equation*}
D ={(n_1+n_2)(n_1+n_2-1)\displaystyle\left({ \sigma_0(k_3)\over\sigma }\right)^2-\left(\sqrt{n_2}\>Y_1-\sqrt{n_1}\>Y_2 \right)^2\over n_1+n_2}-W_1.
\end{equation*}
Similarly, from $p(\overline{X}_1, S_1)\leq k_2$, we get:
\begin{equation}
E_1\leq Y_1 \leq E_2
\end{equation}
with
\begin{equation*}
E_1={\sqrt{n_1}\over\sigma}\left(\dot{\mu}\left(\sigma \sqrt{{W_1 \over n_1-1}}, k_2 \right)-\mu\right),
\end{equation*}
\begin{equation*}
E_2={\sqrt{n_1}\over\sigma}\left(\mu\left(\sigma \sqrt{{W_1 \over n_1-1}}, k_2 \right)-\mu\right)
\end{equation*}
and
\begin{equation}
W_1 \leq F=\displaystyle\left({ \sigma_0(k_2)\over\sigma }\right)^2(n_1-1).
\end{equation}
Setting $W_1=w_1,\>Y_1=y_1,\>Y_2=y_2,\>W_2=w_2$, $S=S(w_1,\>y_1,\>y_2,\>w_2)$, $C_1=C_1(w_1,\>y_1,\>y_2,\>w_2)$, $C_2=C_2(w_1,\>y_1,\>y_2,\>w_2)$, $D=D(w_1,\>y_1,\>y_2)$, $E_1=E_1(w_1)$ and $E_2=E_2(w_1)$, $P(A_2^{u})$ can be written as:
\begin{eqnarray} &&\hspace*{-1.75cm}P(A_2^{u})=\int\limits_0^{F}\biggl(\int\limits_{E_1(w_1)}^{E_2(w_1)}\>\biggl(\int\limits_{-\infty}^{\infty}\>\biggl(\int\limits_0^{D(w_1, y_1,y_2)}H(w_1,y_1,y_2,w_2)\times\nonumber\\
&&\hspace*{.5cm}\times\>g_{n_{2}-1}(w_2)\>dw_2\biggr)\>\Phi'\>(y_2)\>dy_2\biggr)\>\Phi'\>(y_1)\>dy_1\biggr)\>g_{n_{1}-1}(w_1)\>dw_1,
\end{eqnarray}
with\\
\begin{equation*}
H(w_1,y_1,y_2,w_2)=\Phi(C_2(w_1,y_1,y_2,w_2))-\Phi(C_1(w_1,y_1,y_2,w_2)).
\end{equation*}
$P(A_2^{l})$ is obtained by substituting $k_1$ for $k_2$ in $P(A_2^{u})$. Thus, we can state:\\\\
{\bf Theorem 2:} \quad It holds that:
\begin{equation}
L_{\lambda^*_2} (\mu, \sigma)=L_{(n_1, ~k_1)}(\mu, \sigma)+ P(A_2^{u}) -P(A_2^{l}).
\end{equation}
\vspace*{0.5mm}
\begin{center}
4. {\sc The computation of the AM-double sampling plans\\}
\end{center}
\vspace*{2mm}
For a given $p_1$, $p_2$, $\alpha$ and $\beta$, the plan $\lambda^*_2$ is computed in a similar way as $\lambda^*_1$. We use the one-sided approximation $\widetilde{\lambda}_2=\left(\begin {array}{c c c}
n_1 & \tilde{k}_1 & \tilde{k}_2\\ n_2 & \tilde{k}_3
\end {array}\right)$ with
\begin{equation*}
\tilde{k}_1=\Phi\left( {l_1 \over \sqrt{n_1}}\right), ~ \tilde{k}_2=\Phi\left( {l_2 \over \sqrt{n_1}}\right), ~  \tilde{k}_3=\Phi\left( {l_3 \over \sqrt{n_1+n_2}}\right),
\end{equation*}
where $\phi^*_2=\left(\begin {array}{c c c}
n_1 & l_1 & l_2\\ n_2 & l_3
\end {array}\right)$ denotes the AM-double sampling plan in case of an upper tolerance limit $U$ (cf. {\sc Krumbholz} and {\sc Rohr} (2009)). $\phi^*_2$ is determined by
\begin{eqnarray}
&\text{(i)}&L_{\phi_2}(p_1)\geq 1-\alpha\nonumber\\
&\text{(ii)}& L_{\phi_2}(p_2) \leq \beta\\
&\text{(iii)}&N_{max}(\phi_2^*)= \min_{\phi_2 \in Z} N_{max}(\phi_2),\nonumber
\end{eqnarray}
where $Z$ is the set of all double sampling plans $\phi_2$ fulfilling (35)(i) and (ii).
The AM-double sampling plan $\lambda^*_2$ is given
\begin{eqnarray}
&\text{(i)}&\min_{0<\sigma\leq\sigma_0(p)}L_{\lambda^*_2}(\sigma;p_1)\geq 1-\alpha\nonumber\\
&\text{(ii)}& \max_{0<\sigma\leq\sigma_0(p)}L_{\lambda^*_2}(\sigma;p_2) \leq \beta\\
&\text{(iii)}&N_{max}(\phi^*_2)= \min_{\phi_2 \in Z} N_{max}(\phi_2).\nonumber
\end{eqnarray}\\
\underline{Example 1}\\\\
For $~L=1$, $~U=9$, $~p_1=0.01$, $~p_2=0.06$, $~\alpha=\beta=0.1$, we get:
\begin{center}
(i)\hspace*{6mm}$(n,~k)=(36, ~0.02645943143)~~$ and $~~\alpha^*=0.082, ~\beta^*=0.1$,
\end{center}
\begin{center}
(ii)\hspace*{6mm}$\lambda^*_1=\left(\begin {array}{c c c}
26 & 0.017577 & 0.035291\\ 20 & 0.029275
\end {array}\right)~$ and $N_{max}(\lambda^*_1)=32.75439$.
\end{center}
\vspace*{1mm}
\begin{center}
\begin{tabular}{>{\footnotesize}c|>{\footnotesize}c|>{\footnotesize}c|>{\footnotesize}c|>{\footnotesize}c|>{\footnotesize}c}
$\alpha^{**}$ & $\beta^{**}$& $\widetilde{\lambda}_2$ &$N_{max}(\phi^*_2)$ &$\min\limits_{\sigma} L_{\widetilde{\lambda}_2}(\sigma;p_1) $& $ \max\limits_{\sigma}L_{\widetilde{\lambda}_2}(\sigma;p_2) $\\ \hline
0.082& 0.1& $\begin {array}{c c c} 23 & 0.013909 & 0.038143\\
17 & 0.026289\end {array}$ & 30.45689 & 0.8930783818 & 0.0970618822 \\
0.080& 0.1& $\begin {array}{c c c} 23 & 0.013597 & 0.038833\\
17 & 0.026400\end {array}$ & 30.72159 & 0.8955304122 & 0.0969993958 \\
0.078& 0.1& $\begin {array}{c c c} 23 & 0.013993 & 0.038763\\
18 & 0.026496\end {array}$ & 30.99066 & 0.8978607677 & 0.0971362848 \\
0.077& 0.1& $\begin {array}{c c c} 23 & 0.013838 & 0.039100\\
18 & 0.026558\end {array}$ & 31.12727 & 0.8990880486 & 0.0971046378 \\
0.076& 0.1& $\begin {array}{c c c} 23 & 0.013681 & 0.039455\\
18 & 0.026617\end {array}$ & 31.26779 & 0.9003201617 & 0.0970742118 \\
\end{tabular}
\end{center}
\vspace*{4mm}
where $\lambda^*_2=\left(\begin {array}{c c c}
23 & 0.013681 & 0.039455\\ 18 & 0.026617 \end {array}\right)$ with $N_{max}(\lambda^*_2)=31.26778533$.\\\\\\
\underline{Example 2}\\\\
For $~L=1$, $~U=9$, $~p_1=0.01$, $~p_2=0.03$, $~\alpha=\beta=0.1$, we get:
\begin{center}
(i)\hspace*{6mm}$(n,~k)=(115, ~0.0178762881)~~$ and $~~\alpha^*=0.085, ~\beta^*=0.1$,
\end{center}
\begin{center}
(ii)\hspace*{6mm}$\lambda^*_1=\left(\begin {array}{c c c}
81 & 0.014029 & 0.021742\\ 66 & 0.018537
\end {array}\right)~$ and $N_{max}(\lambda^*_1)=103.5432$.
\end{center}
\vspace*{2mm}
\begin{center}
\begin{tabular}{>{\footnotesize}c|>{\footnotesize}c|>{\footnotesize}c|>{\footnotesize}c|>{\footnotesize}c|>{\footnotesize}c}
$\alpha^{**}$ & $\beta^{**}$& $\widetilde{\lambda}_2$ &$N_{max}(\phi^*_2)$ &$\min\limits_{\sigma} L_{\widetilde{\lambda}_2}(\sigma;p_1) $& $ \max\limits_{\sigma}L_{\widetilde{\lambda}_2}(\sigma;p_2) $\\ \hline
0.085& 0.1& $\begin {array}{c c c} 72 & 0.012337 & 0.023495\\
58 & 0.017830\end {array}$ & 98.51959 & 0.8979623972 & 0.0991345737 \\
0.084& 0.1& $\begin {array}{c c c} 72 & 0.012364 & 0.023535\\
59 & 0.017851\end {array}$ & 98.97047 & 0.8990715849 & 0.0991520345 \\
0.083& 0.1& $\begin {array}{c c c} 72 & 0.012385 & 0.023569\\
60 & 0.017875\end {array}$ & 99.43030 & 0.9001786758 & 0.0991672779\\
\end{tabular}
\end{center}
\vspace*{4mm}
where $\lambda^*_2=\left(\begin {array}{c c c}
72 & 0.012385 & 0.023569\\ 60 & 0.017875
\end {array}\right)$ with $N_{max}(\lambda^*_2)=99.43020285$.\\\\\\
{\bf Remark 2:} Numerical investigations indicate:
\begin{enumerate}[(i)]
\item
The AM-double sampling plan $\lambda^*_2$ is more powerful than the AM-double sampling plan $\lambda^*_1$ as it appears that
\begin{equation*}
N_{max}(\lambda^*_2)< N_{max}(\lambda^*_1).
\end{equation*}
\item
Let $\widehat{\lambda}_1$ denote the AM-double sampling plan based on the MVU estimators $\hat{p}_1$ and $\hat{p}_2$ of $p(\mu, \sigma)$. $\hat{p}_1$ and $\hat{p}_2$ are superior over $p^*_1$ and $p^*_2$, respectively, so that
\begin{equation*}
N_{max}(\widehat{\lambda}_1)< N_{max}(\lambda^*_1).
\end{equation*}
For some constellations, it could further be shown that
\begin{equation*}
N_{max}(\lambda^*_2)< N_{max}(\widehat{\lambda}_1)< N_{max}(\lambda^*_1)\quad \text{(See Figure 2)}.
\end{equation*}
\item
The lowest $N_{max}$ among the AM-double sampling plans for a normally distributed quality characteristic with two-sided specification limits and unknown $\sigma$ would feature the plan $\widehat{\lambda}_2$ based on the MVU estimators $\hat{p}_1$ and $\hat{p}_p$ of $p(\mu, \sigma)$, provided that a formula for determining the $\widehat{\lambda}_2$-OC would be found.
\end{enumerate}
\vspace*{-15mm}
\begin{figure}[H]
\begin{center}
   \includegraphics[width=1\columnwidth]{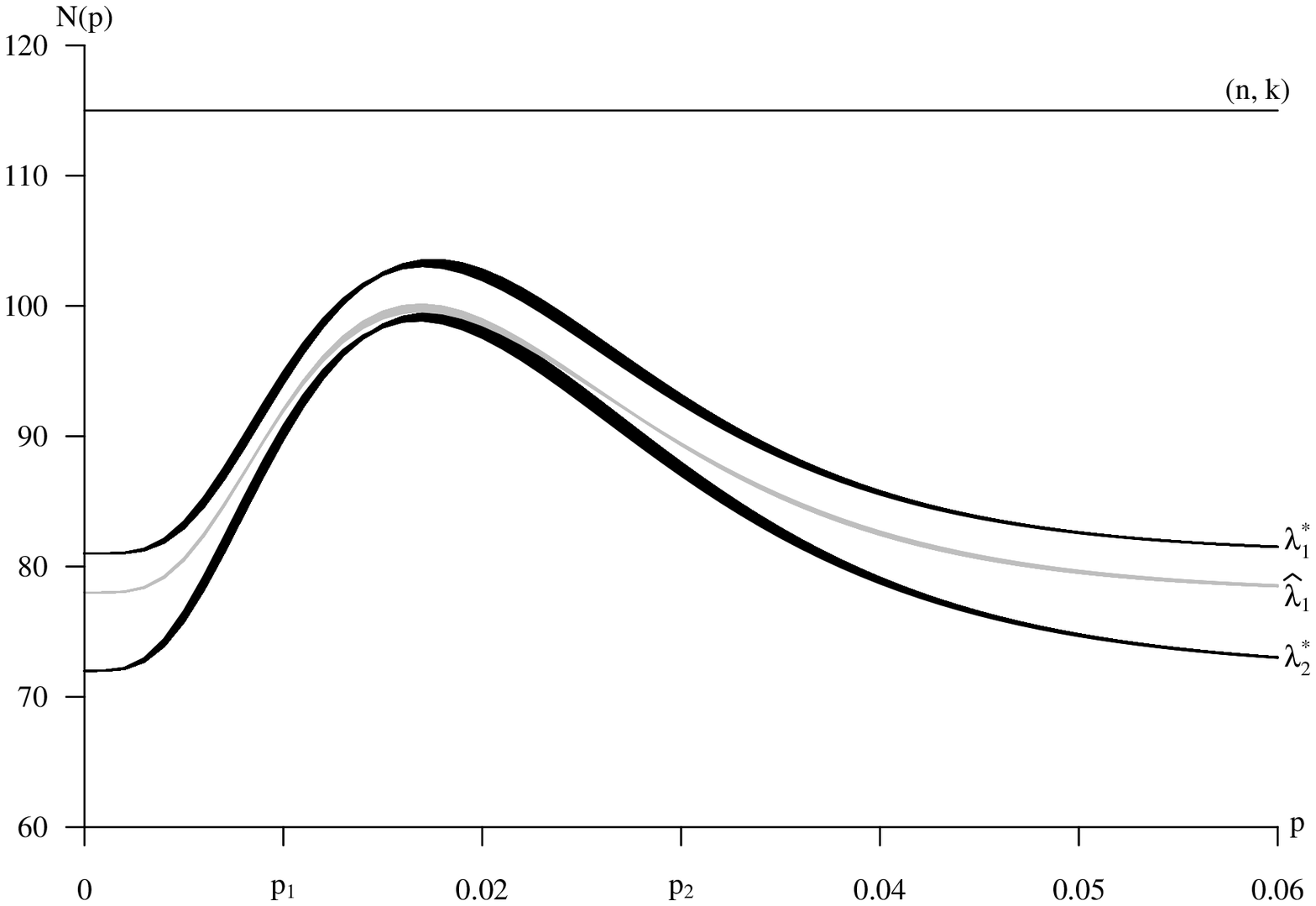}
\end{center}
\end{figure}
\vspace*{-15mm}
Figure 2: ASN bands for $\lambda^*_1$, $\widehat{\lambda}_1$ and $\lambda^*_2$ defined by $p_1=0.01$, $p_2=0.03$ and $\alpha=\beta=0.1$
\begin{center}
{\sc References}
\end{center}
{\sc Bruhn-Suhr, M.}, {\sc Krumbholz, W.} (1990). A new variables sampling plan for normally distributed lots with unknown standard deviation and double specification limits. Statistical Papers 31, 195-207.\\\\
{\sc Krumbholz, W.}, {\sc Rohr, A.} (2006). The operating characteristic of double sampling plans by variables when the standard deviation is unknown. Allgemeines Statistisches Archiv 90, 233-251.\\\\
{\sc Krumbholz, W.}, {\sc Rohr, A.} (2009). Double ASN Minimax sampling plans by variables when the standard deviation is unknown. Advances in Statistical Analysis 93, 281-294.\\\\
{\sc Vangjeli, E.} (2011). ASN-Minimax double sampling plans by variables for two-sided specification limits when $\sigma$ is unknown.\\ \text{http://arxiv.org/PS\_cache/arxiv/pdf/1103/1103.4801v4.pdf}.

\end{document}